# A Conceptual Model for the Organisational Adoption of Information System Security Innovations

Mumtaz Abdul Hameed

Technovation Consulting and Training (Private) Limited. 1/33, Chandhani Magu, Male'.
20-03. Maldives

mumtazabdulhameed@gmail.com

Nalin Asanka Gamagedara Arachchilage

Australian Centre for Cyber Security. University of New South Wales (UNSW

Canberra),

The Australian Defence Force Academy. Australia

nalin.asanka@adfa.edu.au

**Corresponding Author:**

Mumtaz Abdul Hameed

Technovation Consulting and Training (Private) Limited. 1/33, Chandhani Magu, Male'.

20-03. Maldives

mumtazabdulhameed@gmail.com

# A Conceptual Model for the Organisational Adoption of

# Information System Security Innovations

**Abstract**

Information System (IS) Security threats is still a major concern for many organisations. However, most organisations fall short in achieving a successful adoption and implementation of IS security measures. In this paper, we developed a theoretical model for the adoption process of IS Security innovations in organisations. The model was derived by combining four theoretical models of innovation adoption, namely: Diffusion of Innovation theory (DOI), the Technology Acceptance Model (TAM), the Theory of Planned Behaviour (TPB) and the Technology-Organisation-Environment (TOE) framework. The model depicts IS security innovation adoption in organisations, as two decision proceedings. The adoption process from the initiation stage until the acquisition of innovation is considered as a decision made by organisation while the process of innovation assimilation is assumed as a result of the user acceptance of innovation within the organisation. In addition, the model describes the IS Security adoption process progressing in three sequential stages, i.e. pre-adoption, adoption- decision and post-adoption phases. The model also introduces several factors that influence the different stages of IS Security innovation adoption process. This study contributes to IS security literature by proposing an overall model of IS security adoption that includes organisational adoption and user acceptance of innovation in a single illustration. Also, IS security adoption model proposed in this study provides important practical implications for research and practice.

## 1. Introduction

Information and computer resources (hardware, software, database, networks, etc.) collectively be referred as Information System (IS) assets [3] that need to be protected against malicious attacks such as unauthorised access and improper use. Thus, safeguard of IS assets is a widespread concern for individuals and organisations. Research on the preservation of IS assets falls under the theme of IS Security. There are numerous technical measures (software and hardware tools) and non-technical safeguards (physical defences and security procedure) available that provides protection for IS assets. Nevertheless, organisations are still struggling to keep up with threats to their IS assets and security breach incidents that have cost them tens of thousands of dollars in loss [34]. Previous scholarly contributions have constantly argued that the weakest link in any security plan is the computer users themselves [4, 7, 62]. As a matter of fact, computer security education needs to be considered as a means to combat against ISs threats [4, 5, 6, 7].

The main focus of IS security is to deploy strategies to protect and safeguard IS assets from vulnerabilities [3]. However, adoption and implementation of IS security measures in an organisation is a complex process [26]. Besides, adoption of IS security measures by the individuals and organisations is exceptionally low, considering the efforts put in for developing and implementing such systems [37, 57]. Hence, it is critically important to understand what causes the users accept or reject the organisations IS security measures [32]. As far as we can tell from the IS security literature that there is no model that fully explains the IS security adoption in organisations. Nonetheless, research on IS innovation has introduced models, theories and frameworks related to the adoption and implementation of IS innovations in organisations [30]. IS scholars define innovation as an idea, a method, a product, a program or a technology that is new to the adopting unit [17, 30]. Hence, the measures of IS security, undoubtedly, be considered as an IS innovation and the theories based on innovation adoption may obviously be applied in an empirical study on IS security adoption process.

In this research, we aimed to theoretically construct a model for IS security innovation adoption process in organisations, which includes organisational adoption process and the user acceptance of innovation. To this end, we explore the past literature on the stages of innovation adoption, theories of innovation adoption, models of technology acceptance and popular frameworks developed by researchers for organisational adoption, with factors considered to influence IS innovation adoption. This study, then utilised the most suitable concepts and relationships of prominent IS innovation adoption theories and user acceptance models, to explain the process of adoption of IS security innovations in organisations. In addition, this study suggests a number of factors from different context that would either assist or inhibit the process of IS security innovation adoption.

The current study focuses on IS security adoption in organisations. The research makes several contributions to the theory and practice of IS security and innovation adoption research. First, it draws upon and synthesize the rich literature in IS innovation adoption theories and applied it in the context of IS security innovations. The IS security innovation adoption model proposed is based on the theoretical perspective of four innovation adoption theories. The integrated illustration of these models could methodically be used to examine the adoption process and user acceptance of IS security innovations in organisations. Secondly, the proposed IS security adoption model encompasses both the organisational adoption process and user acceptance of innovation. It is evident from the literature that previous scholarly IS security adoption contributions have on no account addressed organisational adoption process and user acceptance of innovation in a single investigation.

Past studies on IS security adoption either examine the processes of adoption of IS security innovation until the acquisition of innovation with no assessment on whether the innovation grows to be part of their regular practice [36, 51]. On the other hand, studies on user acceptance have only examined the behaviour and attitude of individuals accepting an IS security innovation [41, 52]. Combining the organisational adoption process and user acceptance of innovation in a single model allows illustrating the overall adoption process more comprehensively compare to any of the past IS security adoption frameworks. Thirdly, the proposed model has introduced several determinants that could possibly influence IS security adoption in organisations. The suggested association, the study draws between various technological, organisational, environmental, and user acceptance characteristics for IS security adoption provides an abundant opportunity for potential future research. Furthermore, the IS security adoption model proposed in this study provides important practical implications for researchers and organisations.

The remainder of this paper will proceed as follows. First, the theoretical background relating to IS security and different IS security innovations in organisations will be presented. Then, in the 'Methodology' section, we briefly discuss the methods used to identify theoretical elements for the development of model for IS Security innovation adoption in organisations. Following this, in the section 4 'Models and processes of IS Innovation Adoption', we identify the processes of innovation adoption in organisations and the most prominent models use in the IS innovation adoption research. In the 'IS Security Innovation Adoption Model' section, we discuss innovation adoption theories that are most relevant to IS security adoption. In addition, in this section we discuss the synthesis of the IS security innovation adoption model by integrating innovation adoption models. Furthermore, in this section, we presented the proposed model for IS security innovation adoption for organisations. In section 6, 'Determinants of IS security adoption, we suggest a number of factors hypothetically influence adoption of IS security innovation. Conclusion and future research regarding the model were then presented in section 7.

## 2.    Theoretical Background

IS security is, unquestionably, a major concern for most organisations and the risk of computer crimes has become an increasing threat for many companies [14]. As organisations depend more on IS to succeed in their businesses, the management is obliged to invest more on improving their IS reliability [26]. ISs need to be secure if they are to be reliable [35]. Safeguard and the management of Confidentiality, Integrity and Availability (CIA) of information are the most important IS security concerns for an organisation [21].

Security of IS encompass both technical and non-technical concerns for safeguarding IS assets against a variety of threats such as phishing, botnet, virus, worms, Trojans etc. As a safeguard measure, organisations are required to implement policies, practices and technologies that protect against unauthorized access, use, disclosure, disruption, modification or destruction of information [21]. Although there is no such a standard mechanism to completely safeguard all of the IS assets of an organisation, a handful of measures can be put in practice to limit the number of attacks [2]. Recent developments have created new tools and techniques that help organisations effectively secure their IS assets.

A comprehensive range of security measures in the form of physical controls, procedural controls and technical controls would thwart almost all forms of security breaches to ensure CIA of information in an organisation [21]. A wide range physical, procedural and technical security controls can be used to provide security in a number of different ways in an organisation. As a physical control, an organisation may install doors, locks, smoke and fire alarms, fire suppression systems, cameras, flood fences and numerous others. As procedural measures organisation may possibly put in practice data security standards (e.g. use of the IS Management System to keep secure the sensitive company information), the corporate security policies (e.g. restricting access to sensitive personal information to a small number of human resources personnel), password policies (e.g. all system-level passwords must be changed on at least a quarterly basis) and disciplinary procedures (e.g. in case of data breach, the Chief Executive Officer [CEO] will make a decision to inform any external organisation, such as the police or other appropriate regulatory body). To control access to ISs, an organisation may implement technologies such as antivirus, anti-spyware, firewalls, malware programs, filters, network Intrusion Detection Systems (IDS), Intrusion Prevention Systems (IPS), data encryption standards (e.g. Advance Encryption Standards) and authentication-authorization devices. These measures have the potential to reduce and possibly prevent any security risks or vulnerabilities to organisations IS assets.

Any physical, procedural or technical security control put in place in an organisation to protect information and computer resources may possibly be characterized as IS security innovation. As previously outlined, an innovation is the possession of ideas, systems, practices, products or technologies that are new to the adopting organisation. What's more, adoption of innovation is a process that results in the introduction and use of products, processes, or practices that are new to the adopting organisation [18, 30]. Damanpour [17] defines adoption of innovation as the initiation, development and implementation of new initiatives. Hence, implementation and the use of physical, procedural or technical security control may be considered as the adoption of IS security innovation in an organisation.

Correct IS security measures in organisations have long been recognized, however, the empirical research in this area is still at its early stage [8, 9]. Although there are a number of IS security innovations available, an organisation can only benefit if those innovations are adopted and implemented properly. The main hindrance for organisations from attaining a successful implementation of IS security innovation is the lack of appropriate models of IS security adoption. Therefore, this research attempts to examine IS security adoption process in organisations.

## 3.    Research Methodology

The study involved identifying from literature the relevant models and frameworks that could be employed to frame the components that will be used to assess the adoption process of IS security innovations in organisations. Hence, we initially performed a literature search to identify theoretical models utilised in examining adoption and user acceptance of IS innovations. Based on this search result, the study, then identified the most commonly used innovation adoption and user acceptance models. The IS security adoption studies that used IS innovation adoption models in their empirical investigations were then selected. The IS security literature extracted includes studies conducted for both individual and organisational contexts. The most prominent innovation adoption models used in IS security adoption were then drawn together, to synthesize the conceptual model presented in this study. In addition, we extracted the factors from different categories that were examined in the IS security adoption literature.

## 4.    Models and Processes of IS Innovation Adoption

A significant amount of research has been conducted in examining the process and the factors influencing the adoption and user acceptance of innovations in organisations [28, 29, 31]. However, there is no organisational innovation adoption theory that is in existence for researchers to utilis e [30]. Up till now, researchers have been utilising theories and theoretical models from other subjects appropriate to explain the adopter's attitude and innovation adoption behaviour of IS innovation adoption. In addition, innovation adoption research has tailored theories from disciplines such as psychology, sociology and organisational behaviour propose several theoretical models related to the adoption and user acceptance of IS innovations in organisations [30].

The most common theoretical models used to examine adoption and user acceptance of innovation are Diffusion of Innovation Theory (DOI) [49], Perceived Characteristics of Innovation (PCI) [43], Theory of Reasoned Action (TRA) [23], Theory of Planned Behaviour (TPB) [1], Technology Acceptance Model (TAM) [19], Technology Acceptance Model 2 (TAM2) [59], Technology Acceptance Model 3 (TAM3) [58], Technology, Organisation, Environment (TOE) Model [55] and the Unified Theory of Acceptance and Use of Technology (UTAUT) model [60]. Amongst all of these models, DOI, TAM, TRA, TPB and TOE have been widely used in innovation adoption research [30]. DOI, TAM, TRA and TPB are primarily utilised in examining the user behaviour of innovation adoption and TOE framework has widely been exploited in organisational level studies of IS innovation adoption.

Innovation adoption processes in an organisation are considered to be successful only if the innovation is implemented in the organisation and individuals continue to use the innovation over a period of time [25, 30]. Based on this perception, the model presented by Hameed et al. [30] for IT innovation adoption for organisations considered both organisational level analysis and individual level assessment.

Researchers have described the process of adoption of innovation into a number of sequences of stages. According to Hameed et al. [30], the cycle of stages illustrated by different research falls more or less into the initiation, adoption-decision and implementation stage. These three phases of initiation, adoption-decision and implementation are more often referred to as pre-adoption, adoption-decision and post-adoption in the IS literature.

**5.    IS Security Innovation Adoption Model**

This study seeks to develop a conceptual model for IS security adoption that includes the process of adoption and user acceptance of IS security innovations in organisations. A search in literature confirmed that there is hardly any distinct theoretical model with the aim of explaining IS security adoption. IS security research generally utilised IS innovation adoption and user acceptance models [16, 36]. In addition, researchers have applied models from other disciplines such as health belief model to examine user behaviour of IS security adoption [15, 44].

Meanwhile, innovation adoption literature suggests that researchers have been utilising several theories and theoretical models that explain the adopter's attitude and organisational innovation adoption's behaviour to examine different types of innovation such as IS security. As a result, a suitable model or models in the IS field that is general enough may be exploited and perhaps be sufficient to explain IS security adoption in organisations. Indeed, a number of studies have introduced adoption and user acceptance models in the organisational context for various other innovations [28, 31].

### 5.1. Innovation Adoption Models relevant for IS Security

Most of the research on innovation adoption of organisational surrounding conducts their analysis by integrating innovation adoption and user acceptance theories with frameworks that consists of determinants that are relevant to the study context. For example, Teo et al. [53] empirically examined adopters and non-adopters of e-procurement in Singaporean organisations, incorporating two innovation adoption theories and a framework consisting determinant of TOE model. Moreover, Hameed et al. [30] proposed a more general IS innovation adoption model for organisations by combining innovation adoption and user acceptance theories, and major frameworks used in IS innovation studies. Their model is a combination of DOI, TRA, TAM, TPB and a framework that consists of determinants of TOE and CEO characteristics. The model exploited DOI model and the TOE framework with CEO characteristics to illustrate the organisational adoption process until the acquisition of innovation and TRA, TAM and TPB were utilised to construct user acceptance of innovation. Here, TOE framework takes account of the various determinants relevant to IS innovation adoption in organisations.

Consistent with extant research on IS security, we developed the IS Security adoption model by replicating the theories of IS innovation adoption. Based on innovation adoption literature, the study draws together a conceptual model for IS security innovation adoption by integrating multiple theoretical depictions of innovation adoption and user acceptance of IS with popular frameworks. The model is a combination of DOI, TAM, TPB models together with the TOE framework.

### 5.1.1. *Diffusion of Innovation (DOI)*

DOI model introduced by Rogers [49] is the most commonly used theoretical foundations to study innovation adoption. According to Rogers [49], diffusion is a process by which an innovation is communicated through certain channels over a period of time among the members of a group. The DOI theory explains how the individuals or groups adopt innovations and the process involve in their decision towards it. DOI model suggests a number of attributes of innovation that were perceived to assist the diffusion of technological innovation. Rogers [50] suggests that relative advantage, compatibility, complexity, trialability and observability of the innovation plays a key role in an individual's attitudes towards innovation adoption.

The literature shows that the DOI has a solid theoretical basis and the five characteristics suggested in DOI have successfully explained a number of IS adoption behaviours [47]. As for the IS security studies, Lee and Kozar [37] applied DOI to empirically investigate the anti-spyware adoption of computer users of the United State of America.

Nevertheless, Hameed et al. [30] identified two major limitations of the DOI, in its application for organisational innovation adoption. First, the model mainly focuses on the behaviour and attitude of individuals in the adoption of innovation. Another obvious drawback of DOI which Hameed et al. [30] suggested was its inability to address the full innovation adoption process. Hence, the DOI model is inadequate to fully explain IS adoption in organisations. Lee and Kozar [37] argued that DOI model does not clearly explain how an attitude is formed, how it leads to adoption intention and to actual adoption. Nevertheless, the DOI can explain the individual level adoption process in the pre-adoption and adoption-decision stages of innovation adoption. In addition, some researchers have integrated DOI with other theories allowing it to address the interaction between attitude, intention and behaviour [47, 48].

### 5.1.2.   *Technology Acceptance Model (TAM)*

TAM is a persuasive extension or modification of TRA introduced by [19] that aims to explain and predict user acceptance of IS. TAM posits that two cognitive attributes, namely: 'perceived usefulness' and 'perceived ease of use' influences the actual use of IS innovations [19, 20]. According to Davis [19], perceived usefulness is 'the degree to which a person believes that using a particular system would enhance his or her job performance' and perceived ease of use is 'the degree to which a person believes that using a particular system would be free of effort'. Furthermore, TAM articulates that perceived usefulness and perceived ease of use affect a user's attitude towards using an IS and a user's attitude directly relates to a user's intention which eventually determines the system usage of an IS. Among the different models that have been proposed, the TAM appears to be the most widely accepted innovation adoption model among IS researchers [29, 61]. Significantly, TAM has consistently outperformed the TRA in terms of explaining variances across many studies [20, 60]. In addition, TAM has been validated as a powerful and parsimonious framework for explaining user acceptance of IS innovations [19, 20]. Notably, as far as this study is concerned, TAM may possibly be utilized to investigate IS security implementation. Meanwhile, in a research to examine the factors that influence employee acceptance of IS security measures, Jones et al. [32] extended the TAM.

While IS researchers have investigated and replicated TAM, and agreed that it is suitable in predicting the individual acceptance of IS innovations, the TAM's fundamental constructs do not fully unveil the influence of contextual factors that may affect the users' acceptance of IS innovation [30]. Further, Legris et al. [38] argued that TAM needed to be integrated into a larger model that includes technological and social factors to enhance its analytical capability to predict innovation acceptance of IS innovations. One of the shortcoming of TAM is that it only considers the individual level acceptance and neglects group level aspects of decision making [30]. Another limitation of TAM is that, it theorises user's adoption decision based purely on voluntary situations, neglecting users' judgement influenced by their peers or in response to social pressure. To overcome these limitations, researchers have incorporated TAM with other IS adoption models to examine organisational adoption process [48, 53].

### 5.1.3.    *Theory of Planned Behaviour (TPB)*

TRA envisaged that the individual acceptance behaviour is purely voluntary, however, many decisions by an individual does not appear fully volitional. Hence, to address the non-voluntary actions of user acceptance behaviour, Ajzen [1] revised TRA by adding a new component to develop an improved replica known as the TPB. According to TPB, human action is guided by three kinds of thoughts: (1) behavioural beliefs - one's opinions about the likely consequence of the behaviour, (2) normative beliefs - one's perception about the normative expectation of others and (3) control beliefs - one's judgments about the presence of factors that may facilitate or impede the performance of the behaviour. TPB's behavioural belief intends to produce a favourable or unfavourable attitude towards the behaviour and normative belief is the result of perceived social pressure or subjective norm to perform the action. In addition, TPB's control beliefs, namely Perceived Behavioural Control (PBC), foresees the decisions regarding the absence or presence of factors that might facilitate or impede the performance of the behaviour [1].

TPB proposes that a combination of attitude toward the behaviour, subjective norm, and PBC may lead to the formation of a behavioural intention to perform the behaviour. TPB suggests that PBC affects behaviour directly or indirectly through behavioural intention. Ajzen [1] argued that in conditions where behavioural intention has minimal effect on the actual behaviour, PBC alone can be used to predict the behaviour. Armitage and Conner [10] showed that PBC was a significant factor in the prediction of behavioural intention and actual behaviour in TPB, regardless of the effects of attitude toward the behaviour and subjective norm.

In a meta-analytic review of TPB, Armitage and Connor [10] concluded that the theory is an effective model to validate user acceptance of innovation and the three antecedents of TPB model directly and indirectly predict individual behaviour for a number of innovations. TPB has been used in numerous IS innovation adoption contexts to predict and explain individual behavioural intentions as well as the actual use of innovation [12, 60].

TPB model has also been used to examine IS security innovation adoption studies. For example, Lee and Kozar [36] used TPB model to identify the factors influencing the user adoption of anti-spyware systems. The research examines the influence of three constructs of TPB i.e. attitude, social influence and Perceived Behavioural Control (PBC) for anti-spyware adoption of individuals. Similarly, in a review to observe the user behaviour to conscious care behaviour in the domain of information security, Safa et al. [51] utilised TPB model. Lee and Kozar [37] applied DOI and TPB model for an empirical investigation of anti-spyware adoption of computer users. The study investigates the attributes of DOI and TPB for user's anti-spyware adoption intention.

Like all other innovation adoption models discussed above, TPB has certain limitations, which needs to be considered when applied in the innovation adoption research. Albeit TPB considers normative influences, it still does not take into account the influence of environmental or organisational factors that may influence the innovation adoption. As a result, researchers extend TPB by combining its constructs with the components from other contextual frameworks such as TOE.

### 5.1.4. TOE Model

Tornatzky and Fleischer [55] synthesized a structure for organisational innovation adoption based on the Contingency Theory of Organisations. The framework identified determinants of technology, organisation and environment as three dimensions of a firm that affects organisational adoption. Hence, the framework was named as "TOE" framework. In this framework, the technological context relates to both internal and external technologies available to an organisation. Its main focus is on how the existing technologies within the organisation as well as the available innovations external to the firm influences the innovation adoption process. The organisational context describes the impact of the characteristics of firms on innovation adoption process. Common organisation characteristics include firm size, degree of centralization and formalization, the complexity of its management structure, the quality of its human resources, and the amount of slack resources available internally. The external environmental context is the arena, in which an organisation conducts its business [55]. This includes the industry, competitors, regulations, and relationships with the government.

TOE framework unravels the limitations of other innovation adoption and user acceptance models to predict innovation adoption in organisations. TOE model describes the impact of specific attributes in the three contextual domains of organisations towards the innovation adoption process. TOE framework can be combined with other theories to better explain IS innovation adoption in organisations.

## 5.2. *Synthesizing IS Security Innovation Adoption Model*

The model for the adoption of IS security innovation in organisations we presented in this study consist of a combination of innovation adoption and user acceptance theories jointly with contextual frameworks of IT innovation adoption. It is evident from the literature that previous scholarly IS security adoption contributions have on no account addressed organisational adoption process and user acceptance of innovation in a single investigation. By and large, IS security adoption studies have examined individual attitudes and behaviour towards innovation [32, 37, 51]. Research on IS security rarely considers the adoption process at the organisational level.

### 5.2.1. *IS Security Innovation Adoption Process*

Establishing the views of IS innovation literature and consistent with the model presented by Hameed et al. [30], the model we presented in this study discusses IS security innovation adoption for organisations as a two level adoption process, an organisational level analysis and individual or user level evaluation. The developments the organisations undergo from the introduction of the IS security innovation until the actual acquisition of the innovation is regarded as an organisational level adoption process. Following that, the user acceptance of innovation and the continuous use of the innovation as an IS security application within the organisation is classified as an individual level adoption process.

To concur with the basics of much of the previous research in IS innovation adoption [30, 45, 50] this study considered IS security innovation adoption process in organisations as a three stage process of pre-adoption, adoption-decision and post-adoption. The study deems that the pre-adoption stage consisting of activities related to recognizing a need, acquiring knowledge or awareness, forming an attitude towards the innovation and proposing innovation for adoption [25, 50]. The adoption-decision stage described by Hameed et al. [30] reflects the decision to accept the idea and evaluates the proposed ideas from a technical, financial and strategic perspective, together with the allocation of resources for its acquisition and implementation. The study also considered the post-adoption stage, which involves the acquisition of innovation, preparing the organisation for the use of the innovation, performing a trial for confirmation of innovation, acceptance of the innovation by users and continued actual use of the innovation [50].

### 5.2.2. *Integrating Innovation Adoption Models*

The study synthesized the conceptual model for IS security innovation adoption in organisations by integrating four theoretical replicas of innovation adoption and user acceptance of IS. The model is an integrated combination of DOI, TAM, TPB models together with the TOE framework. DOI is the most generally accepted model for identifying the main characteristics of IS innovation adoption [30, 47, 54]. However, the model cannot be used to explain full IS security innovation adoption process as it does not include the post-adoption behaviour of the innovation adoption. In addition, DOI only explicates the individual level adoption process which inhibits the model to utilise in explaining IS security innovation adoption in organisations. Combining DOI with TAM and TPB helps us to derive a model that reflects pre-adoption, adoption-decision and post-adoption stages of IS security innovation adoption. TAM and TPB have been used in empirical investigations to predict and explain user acceptance of IS innovation [29]. TPB complements TAM's constructs at the same time TPB explanatory and predictive power enhances further by integrating with TAM [11]. Furthermore, TAM combined with TPB constructs would allow predicting user acceptance of innovations for both volitional and non-volitional conditions [30]. DOI, TAM and TPB models have been successfully exploited and effectively been used in explaining and predicting either the adoption or user acceptance of IS innovations by individuals [30]. In order for our IS security adoption model to address organisational level adoption process, integrative illustration of DOI, TAM and TPB have to be combined with a contextual framework. The TOE model has been extensively adapted for the studies of IS adoption in organisations [47, 48, 53]. Thus, an integrative model consisting of DOI, TAM, TPB and TOE would fully explain IS security innovation adoption in organisations.

Use of DOI and TOE in the proposed model could successfully explicate the adoption process at organisation perspective. In light of the technology, organisation and environment attributes that facilitate the adoption, both DOI and TOE competently elucidate pre-adoption and adoption-decision stages of IS security adoption in organisations. The proposed model uses the constructs of TAM and TPB to account for the user acceptance of IS security innovation. Hence, the user acceptance attributes of TAM and TPB affects the IS security adoption process at the post-adoption stage. TPB has been a particularly useful model to predict user acceptance of IS innovations in organisations, where the use of IS is not entirely under the volitional control of the user.

### 5.3. *Proposed Model of IS Security Innovation in Organisation*

Figure 1 illustrates our proposed conceptual model for the IS security innovation adoption in organisations.

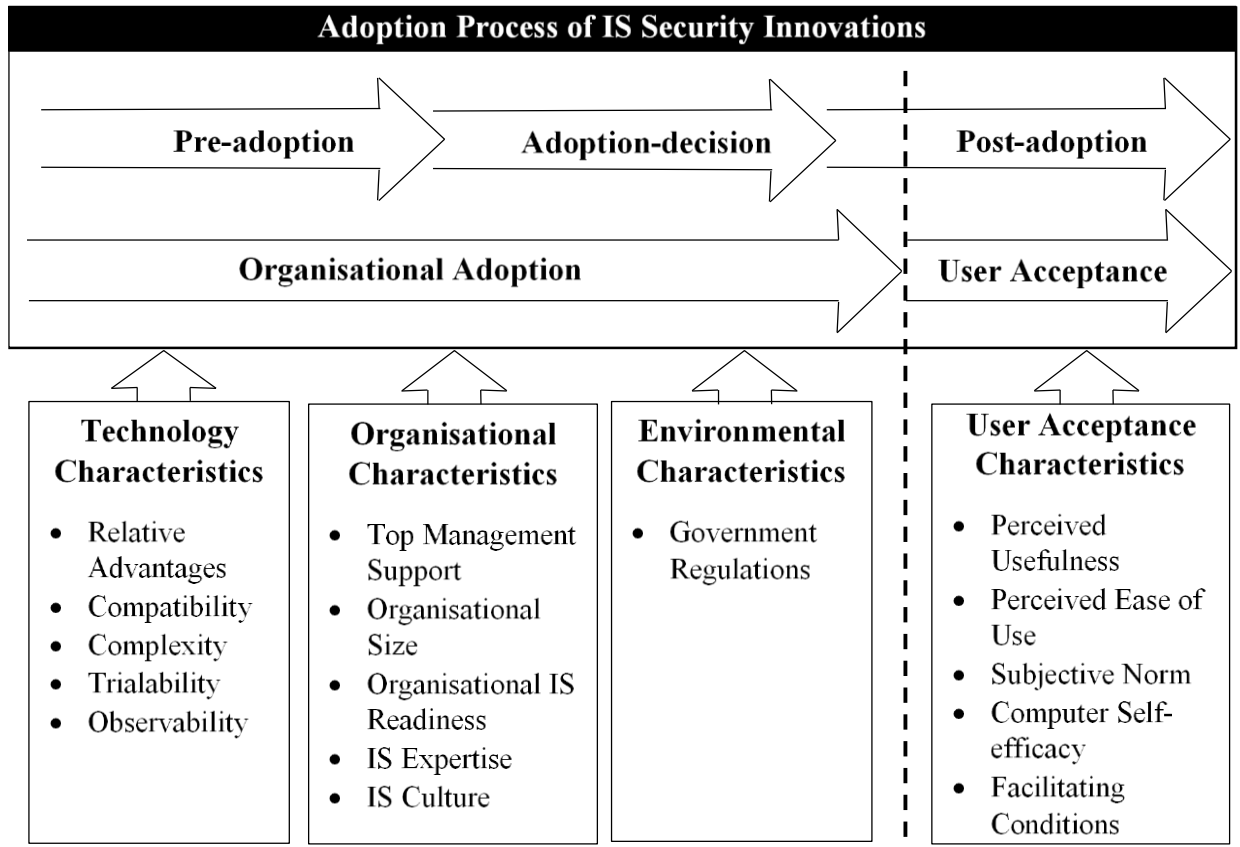

***Figure 1: Model for the Adoption of Information System Security Innovation in Organisations***

## 6.    Determinants of IS Security Innovation Adoption

For the IS security adoption model shown in the Figure 1, we considered technology, organisation, environment and user acceptance attributes that were examined in the past IS innovation adoption literature. In addition, each of the factors included in proposed model have been examined and found to have a significant influence on IS security innovation adoption research.

### *6.1.    Technology Characteristics*

Technological context of TOE model outlines a number of attributes of technologies inside the organisation and the innovations available outside the firm. The importance of technology attributes for the adoption and implementation of IS and the perception of innovations influencing the pre-adoption and adoption-decisions have been documented in the IS literature [49]. Specific characteristics of innovation are examined as factors that explain innovation adoption in organisations. DOI theory provides a set of innovation attributes that may affect the adoption decision [50].

For this study, we considered the relative advantage, compatibility, complexity, trialability and observability in terms of technology characteristics. Lee and Kozar [36] and Lee and Kozar [37] examined anti spyware software adoption and found that relative advantage, compatibility, trialability and visibility (observability) as important determinants. Salleh et al. [52] suggests that perceived complexity and perceived compatibility of innovation are important attributes of big data security solutions adoption.

### 6.1.1. *Relative Advantage*

Relative Advantage is the degree to which the innovation is perceived as better than the idea it supersedes [28]. Relative advantage refers to the degree to which the innovation is more productive, costs saving, less maintenance, efficient compared to the existing practices [50]. Indeed, relative advantage is one of the key determinant that would influence a person or an organisation to adopt an IS innovation [28, 43]. Hence, we suggest that organisations are more likely to adopt IS security innovation when they perceive that it is a valuable and effective means of protecting their IS assets.

### 6.1.2. *Compatibility*

Compatibility is the degree to which an innovation is perceived as being consistent with the existing values, past experiences and needs of the users [28, 50]. An innovation must be considered generally acceptable if it is to be implemented successfully. The more compatible the new innovation is with the existing processes and systems, the more easily the innovation gets implemented and integrated into the organisation [28]. According to Lee and Kozar [36], if the IS security innovation is compatible with the current system, the less resistance the organisation will experience accepting the innovation. Hence, compatibility of an IS security innovation is positively related to adoption and implementation within the organisation.

### 6.1.3. *Complexity*

Complexity is the degree to which the innovation is perceived as difficult to understand and use [50]. Innovations that are simple to comprehend are more likely to be adopted by organisations. The complexity of an innovation is expected to influence negatively for the adoption of IS innovations in organisations [28, 56]. Hence, less complex IS security innovations are believed to adopt faster and propagates a smooth implementation process, thereby achieving the efficiency anticipated.

### 6.1.4. *Trialability*

Rogers [50] defines "trialibility" as the degree to which the innovation may be experimented with. Hence, IS innovations are likely to be adopted if they can be tried out on a temporary basis. Being able to try an innovation before adoption reduces the uncertainty of potential adopters [56]. Trialability is important in the initiation stages of innovation adoption, however, its implication will affect the usage of the innovation. The literature suggests a positive relationship between trialability and innovation adoption [28, 50]. In this model, we assume that the better exposure one gets to a particular IS security innovation, the more likely that it will be adopted and used in an organisation.

### 6.1.5.    *Observability*

Observability is the degree to which the results and the advantages of an innovation are visible to others [50]. Hence, observability is sometimes referred to as "visibility." Lee and Kozar [37] investigated the adoption of anti-spyware software and found that observability has a significant impact on the probability of adoption. The more visible or observable the usage and the outcome of an IS security innovation, the more likely the innovation will be adopted and implemented in an organisation. Observability is expected to have a positive relationship with adoption of innovations [50].

### 6.2.    *Organisational Characteristics*

The organisational context of TOE model has been the most frequently examined attributes in adoption of IS innovations in organisations. The organisational attributes, addresses the facilitating and inhibiting factors in the area of operations in a firm. Researchers have advocated the primary importance of organisational determinants compared to other contexts as predictors for innovation adoption [17, 31]. Organisations are adopting innovations in response to an external demand or to achieve an advantage of an environmental opportunity [27]. For this study, we propose top management support, organisational size, organisational IS readiness, IS expertise and IS culture as organisational attributes that influence IS security adoption.

To verify IS conscious care behaviour formation in organisations, Safa et al. [51] found IS awareness, IS experience and organisation policy as important factors. IS awareness and IS experience described by Safa et al. [51] is often described as a single attribute in the IS literature, namely: IS expertise [31]. Hence, for the proposed model of IS security adoption, we refer IS expertise that takes account of the organisational awareness and experience of the innovation. Li [41] verified that the size of the organisations impacts on online security performance in organisations. Salleh et al. [52] utilised the TOE framework to explore top management support, IS culture and organisational learning culture as security determinants of big data solutions. According to Leidner and Kayworth [39], IS culture is a variable that explains how social groups interacts with IS in the development, adoption, use and its management. Hence, IS policy suggested by Safa et al. [51] and IS learning culture described by Salleh et al. [52] could simply be classified under IS culture. In our proposed model, we included IS culture as a determinant that accounts the perception of IS policy and IS learning culture. Lee and Kozar [36] and Lee and Kozar [37] found that computing capacity of the organisation significantly influence anti-spyware systems adoption. In the IS literature, computing capacity is termed as IS readiness and we adopt this terminology in our model.

### 6.2.1. *Top Management Support*

A recurring, organisational factor studied by IS researchers is top management support. Top management support is one of the consistently found and a highly critical factor that influence IS implementation [54]. It is commonly believed that top management support plays a vital role in all stages of adoption of IS [31]. There is also evidence in the innovation literature that suggests top management support is positively related to the adoption of new technologies in organisations [56]. For IS security adoption, top management's role in allocating the required resource to safeguard IS asserts and to provide a supportive climate in the user acceptance of innovations are vital.

### 6.2.2. *Organisational Size*

Organisational size has been the most frequently examined factor in the study of organisational innovation adoption [47]. Organisational size is the most important factor influencing IS adoption, since, the size of an organisation determines other organisational aspects, particularly slack resources, decision making and organisational structure [50]. As the size of the organisation gives a good indication of the slack resources available, we suggest that organisational size has a positive influence on the adoption of IS security innovation.

### 6.2.3. *Organisational IS Readiness*

Organisational IS readiness is defined as the degree to which an organisation has the knowledge, resources, commitment and governance to adopt IS innovations [31]. Adoption of IS has often been positively associated with organisational IS readiness. IS infrastructure and computing resources are essential to successfully implement and gain advantages from any IS adoption [46]. Prior studies revealed a positive association between the existence of IS readiness and adoption of IS [31]. We suggest that the existence of IS infrastructure and the availability of financial and technological resources within an organisation influence the adoption of IS security innovations.

### 6.2.4. *IS Expertise*

In an organisation, knowledge of IT is a major factor in the adoption of new technologies [22]. An organisation with existing knowledge of new innovation makes adoption effortless and retains knowledge of innovation adoption [40]. For the study, firm's possession of the awareness of IS security and expertise of IS security threats within an organisation helps the adoption of IS security innovations.

### 6.2.5. *IS Culture*

Introduction of a new IS fundamentally changes the way the organisation solves problems and this process results in the creation of a new IS culture. Organisational culture has also been shown to play a significant role in IS management policies [13]. Organisational culture can support IS adoption and can thus be a critical success factor for the development and implementation of IS innovations.

Furnell and Thomson [24] investigated IS culture and found to have a positive effect on IS security adoption. We propose that organisational beliefs and values regarding IS security and the existing IS security policies play an important role in the IS behaviour of an organisation.

### 6.3. Environmental Characteristics

IS has not only been used for internal needs; instead, organisations often communicate with customers, suppliers and other trading partners [27]. Hence, environmental factors are increasingly being studied in innovation adoption studies. The recommended attribute of this study in terms of environmental context is government regulation [27, 33]. Li [41] verified that the existence of government regulation significantly influences online security adoption in organisations.

### 6.3.1. Government Regulations

The regulatory environment and governmental institutions enforce policies on taxes, trade, investment, patents, product liability, consumer protection and human resources that may have a powerful effect on technology adoption. Numerous researchers have highlighted the role of government regulations on the adoption decision of IS innovation [27]. This study suggests that government's support and regulatory policies may have a huge impact on IS security adoption and implementation.

### 6.4. User Acceptance Characteristics

Constructs of TAM and TPB contribute most towards user acceptance attributes. The two attributes of TAM, perceived usefulness and perceived ease of use were key determinants of user IS acceptance [29]. The constructs of TPB namely: attitude, subjective norm and PBC were also key determinants of user acceptance of IS innovations [10]. Furthermore, two sub-constructs of PBC, computer self-efficacy and facilitating conditions which predicts the non-volitional behaviours were also found to be important characteristics [29].

Lee and Kozar [36] found that user acceptance attributes of attitude, social influence and PBC significantly influence anti-spyware systems adoption. Lee and Kozar [37] examined anti-spyware software adoption and found that user acceptance attributes of attitude, subjective norm and self-efficacy are important determinants. Jones et al. [32] examined TAM attributes and found perceived usefulness; perceived ease of use and subjective norm had a significant impact on user intention to use IS security measures. To verify IS conscious care behaviour formation in organisations, Safa et al. [51] found an important relationship with user attitude, subjective norm and self-efficacy.

As our proposed model uses the constructs of TAM and TPB to account for the user acceptance of IS security innovation, the factors included were perceived usefulness, perceived ease of use, subjective norm, computer self-efficacy and facilitating condition. Venkatesh and Davis [59] and Hameed and Counsell [29] also considered these five characteristics to determine user acceptance of IS innovation.

### 6.4.1. *Perceived Usefulness*

Perceived usefulness is defined as "the degree to which a person believes that using a particular system would enhance his or her job performance [19]. Perceived usefulness is a major determinant of intention to use the innovation and were found to have a direct effect on the usage behaviour. Much of the previous research has investigated TAM, and confirmed that perceived usefulness is the strongest predictor of an individual's intention to use an innovation [59, 60]. In our proposed model, we define perceived usefulness as the degree which as an individual believes that the use of IS security innovation will virtually secure his or her IS assets.

### 6.4.2. Perceived Ease of Use

Perceived ease of use is "the degree to which a person believes that using a particular system would be free of effort [19]. TAM suggests that perceived ease of use has a significant influence on perceived usefulness, behavioural attitude, intention, and actual use of an innovation [20]. In this research, we define perceived ease of use as the extent to which an individual perceives that the interaction with the IS security innovations is effortless.

### 6.4.3. Subjective Norm

Fishbein and Ajzen [23] described subjective norm as the pressure enforce on individuals by people or organisations important to them, to perform or not to perform a particular behaviour. For our proposed model, subjective norm is the social pressure on the employee of an organisation by the management, supervisor, head of department and colleagues to accept or reject an IS security innovation. This pressure affects the employee's acceptance decision and the use of IS security innovations.

### 6.4.4. Computer Self-efficacy

Computer self-efficacy is defined as an individual's self-confidence in one's ability to perform a behaviour [29]. In the context of IS security adoption, the dimension of self-efficacy is defined as an individual's beliefs about their ability to proficiently implement preventive security behaviours and use preventive security tools [62]. In the information security adoption literature, self-efficacy has been found to be significantly associated with both intentions [42] and use [44] of IS security innovations. Hence, it is expected that higher levels of computer self-efficacy will cause higher levels of PBC, thus with increased intentions to use IS security innovation.

### 6.4.5. *Facilitating Condition*

Facilitating conditions is the degree to which an individual believes that an organisational and technical infrastructure exists to support the use of the innovation [60]. IS security requires preventive software and tools in addition to technical supports to the user [62]. This study suggests that the more support services available to the users, the better chance that the users accept and trust the IS security innovation.

## 7. Conclusions and Future Research

Effective adoption of IS security innovation is critical in protecting organisation's IS assets from malicious attacks. In this study, we developed and proposed a model for the process of IS security innovation adoption in organisations. The study considered the IS security innovation adoption process to be successful only if the innovation is accepted and integrated into the organisation and the individual users continue using the innovation. The study described IS innovation adoption process as a two level decision proceeding: an organisational level adoption judgement followed by the verdict of the users regarding the innovation. Thus, the model exemplifies a two levels of evaluation, i.e. from the initiation stage until the acquisition of innovation was assessed as organisational process and the process of user acceptance of the innovation is analysed in terms of the behaviour of the individuals within the organisation. Furthermore, the proposed model portrayed the IS security adoption process, as progressing in three distinct phases, from pre-adoption through adoption-decision and then post-adoption stages.

The basis of the model is derived by replicating theories and models used in the studies of innovation adoption and user acceptance of IS innovation. The study integrated perspectives from DOI, TAM, TPB and TOE to depict IS security innovation adoption process in organisations. The model exploited DOI model and TOE framework to characterize the organisational adoption process until the acquisition of innovation and the constructs of TAM and TPB explains the user acceptance of IS security innovation. The model also included several factors from different contexts that is perceived to impact IS security innovation adoption in organisations. The study suggests that relative advantage, compatibility, complexity, trialability and observability of the innovation influences the IS security adoption in an organisation. In terms of organisation characteristics the study proposes top management support, the size of the firm, IS readiness, IS expertise and IS culture to impact IS security innovation adoption process. In addition, the study advocates that government regulations are key to the adoption decision of IS security adoption in organisations. The study suggests that perceived usefulness, perceived ease of use, subjective norm, computer self-efficacy and facilitating conditions are factors that enables users to accept IS security innovations.

The proposed model presented have considerable significance in understanding the process involved in the adoption of IS security innovation in organisations. Also, it allows to highlight the key steps to navigate to achieve a successful adoption of IS security innovations. Equally, the study provides researchers and practitioners with a set of factors that affect the adoption of IS security in organisations. It serves as a guideline for practitioners to identify and address the facilitating and inhibiting issues in the context of technology, organisation, environment and user acceptance attributes in the process of IS security adoption. Managers need to consider these issues when embarking on IS security adoption in their organisations.

The contribution of the study includes an enhancement of our understanding of IS security adoption and implementation process in organisations. It draws upon and merged from the rich literature in IS innovation adoption theories and applies it in the context of IS security where, it has seldom been examined. To overcome the shortcomings of individual IS innovation adoption models such as the DOI and TAM; the proposed model combined a number of innovation adoption models. Combining different innovation adoption models allows the individual model to complement each other, strengthening the analytical ability of the proposed model. Another important contribution of this research is that the proposed IS security model considers both the organisational adoption process and the user acceptance of innovation in a single illustration. Incorporating the organisational adoption process and the user acceptance of innovation in a single representation allows to explain IS security innovation adoption process more thoroughly. In addition, the proposed model introduces several determinants from technology, organisation, environment and user acceptance characteristics that may influence IS security adoption in organisations.

The IS security adoption model proposed in this study provides important implications for practice as well as for further research. This study has a number of implications for managers and IS researchers. Managers can draw up this model and assess the condition of the IS security adoption process and possible factors that would lead to a successful adoption of IS security innovations in their organisations. In addition, managers can utilise the model to plan and prepare for the adoption process and establish smooth conditions for the user acceptance in the IS security implementation process. IT practitioners may utilise this model to investigate the factors influencing the adoption of IS security innovations in various demographic settings; the model could be tested by organisations from different sectors and different countries.

The study has some limitations that need to be considered when interpreting the results. As this is a theoretical model, unavailability of any empirical insights of the model limits us to drawing causal implications of the findings. Another limitation is that this research obtained technology, organisation, environment and user acceptance attributes from a small number of studies. This could result in a narrow scope that does not adequately capture all determinants relating to IS security adoption.

In terms of future research, the proposed model may be empirically tested to verify the effectiveness in predicting IS security adoption in organisations. Another obvious addition to this study would be to experimentally examine the impact of the attributes identified in the proposed model. As the model gives no indication of the significance of each factor for different stages of IS security adoption, the researchers could extend this study by analysing the interaction between different characteristics at different stages of the IS security adoption process. Furthermore, future research may refine the relationships using an empirical investigation to enable researchers to establish causal relationships.